\documentclass[letterpaper, 10 pt, conference]{ieeetran}  

\IEEEoverridecommandlockouts                              

\usepackage{graphics} 
\usepackage{epsfig} 
\usepackage{amsmath} 
\usepackage{amssymb}  
\usepackage{amsfonts}                                
\usepackage{enumerate}
\usepackage{psfrag}

\def\qedp{\hspace*{\fill}~{\tiny $\blacksquare$}}
\def\salt{\vskip 0.2 true cm}

\newtheorem{theorem}{Theorem}
\newtheorem{itlemma}{Lemma}
\newtheorem{itdefinition}{Definition}
\newtheorem{itproposition}{Proposition}
\newtheorem{itfact}{Fact}
\newtheorem{itremark}{Remark}
\newtheorem{itassumption}{Assumption}
\newtheorem{itcorollary}{Corollary}
\newtheorem{itexample}{Example}
\newenvironment{definition}{\begin{itdefinition}\rm}{\end{itdefinition}}

\newenvironment{proposition}{\begin{itproposition}\rm}{\end{itproposition}}
\newenvironment{remark}{\begin{itremark}\rm}{\end{itremark}}

\newenvironment{lemma}{\begin{itlemma}\rm}{\end{itlemma}}

\title{Switching Control for Parameter Identifiability of Uncertain Systems 
}

\author{Giorgio Battistelli  and Pietro Tesi
\thanks{G. Battistelli is with the
        University of Florence, Dipartimento di Ingegneria dell’Informazione (DINFO), Via di Santa Marta 3, 
        50139 Firenze, Italy (e-mail: giorgio.battistelli@unifi.it).  P. Tesi is with University of Groningen,  
        ENgineering and TEchnology institute Groningen (ENTEG), 
        Faculty of Mathematics and Natural Sciences, Nijenborgh 4, 9747 AG Groningen,
        The Netherlands (e-mail: p.tesi@rug.nl).
       }
}

\begin{document}

\maketitle
\thispagestyle{empty}
\pagestyle{empty}

\begin{abstract}
This paper considers the problem of identifying the parameters of an uncertain 
linear system by means of feedback control. The problem 
is approached by considering time-varying controllers.
It is shown that even when the uncertainty set is not finite, 
parameter identifiability can be generically ensured by 
switching among a finite number of linear time-invariant controllers. 
The results are shown to have several implications, ranging from 
fault detection and isolation to adaptive and supervisory control.
Practical aspects of the problem are also discussed in details.
%
\end{abstract}

\section{introduction}

Identifying the parameters of an uncertain 
system from input-output data is  
a problem of long-standing fundamental interest in control engineering. 
This problem is often referred to as the problem of \emph{parameter identifiability} 
\cite{Willems,glover}. 
This paper considers the identifiability problem 
with respect to uncertain linear systems where the uncertainty 
set consists of a known bounded set possibly containing 
a continuum of parameters. For this class of systems,
we address the problem of ensuring the identifiability 
of the unknown parameters of the system by means of feedback controllers.

The motivations for studying this problem are immense. For instance, parameter identifiability of a feedback loop can be of interest in
the context of {\em fault detection/isolation} for systems subjects to failures, in order to make it possible to promptly detect any departure from the nominal
behavior and to precisely identify the parameter variation \cite{kinnaert,cocquempot}.
Another application is {\em control reconfiguration} wherein the objective is that
of replacing the active controller (typically designed in order to ensure robust stability in all the uncertainty region) with a different one providing
enhanced (possibly optimized) performance \cite{liberzon,steffen}. Finally, on-line estimation of the uncertain parameters under feedback can be exploited when dealing with systems which naturally exhibit multiple operating conditions  for constructing {\em adaptive control} laws. In fact, many existing adaptive control techniques rely on the idea of \emph{certainty equivalence}, 
which amounts to applying at each instant  of time the controller designed for 
the model that best fits the available data \cite{Astrom,Ioannou}. 

In this paper, the problem of parameter identifiability 
is approached by searching for feedback control laws 
under which closed-loop behaviors obtained 
with different system parameters can be distinguished one from another.
To this end, we introduce a notion of \emph{discerning control}.
Parameter identifiability is then defined precisely in terms of 
discerning control.

In principle, parameter identifiability 
under feedback can always be ensured by means 
of a \emph{probing signal} injected into the plant as an additive perturbation input, 
superimposed to the control variable \cite{Astrom,Ioannou}. Nevertheless, in many contexts, such a solution should be avoided due to the inherent drawback of leading the feedback loop away from the desired behavior,
thus destroying regulation properties. This is especially true when the behavior of the feedback loop has to be monitored continuously as in the contexts of fault/detection isolation and adaptive control.
Then, a natural question arises on whether or not it is possible to guarantee parameter identifiability directly by means of a feedback control law, possibly designed also to satisfy
other control objectives (\emph{e.g.}, stability in nominal operating conditions). 
An affirmative answer to this question was given in \cite{Ba13,BaBaTe14} for the special case
of switching linear systems, \emph{i.e.}, when the uncertain parameters can take on only a finite number of possible values. Specifically, in \cite{Ba13,BaBaTe14}, it is shown that, under quite mild assumptions, for switching linear systems almost all linear time-invariant (LTI) controllers are discerning. 

In general, an analogous result cannot be established in case of 
continuously parameterized systems, \emph{i.e.}, when the uncertainty set is not finite.
The reason is inherently tied to the fact that LTI controllers do not generally 
provide a sufficient level of excitation to the loop \cite[Chapter 2]{Astrom}.
The problem of loss of identifiability due to feedback can be overcome 
by means of time-varying controllers, and one possibility 
is given by switching control \cite{MOPA,MOPA2}. 
In this paper, we exploit this property
to show that parameter identifiability can in fact be \emph{generically} ensured by 
switching among a finite number of LTI controllers (hereafter called modes), provided that the number of different controller modes is sufficiently large. 
Specifically, an upper bound on the number of controller modes needed for parameter identifiability is given in terms of the dimension of the uncertainty set.
Moreover, we show that the result remains
true even if we restrict the controller modes to be of a given fixed order
and to satisfy certain stability requirements. 
The latter result is perhaps surprising as it indicates that
the seemingly conflicting goals of ensuring parameter identifiability as well as a satisfactory behavior of the feedback system (at least under nominal conditions) 
can be simultaneously accomplished by means of switching control. 

As a further contribution, we analyze the properties of 
least-squares parameter estimation in connection with the use of discerning controllers.
Specifically, in order to ensure the practical applicability of the estimation technique, we focus on a {\em multi-model} approach wherein the estimate is selected among 
a finite number of possible values of the parameter vector (obtained by suitably sampling the uncertainty set).
In this context, a bound on the worst-case parameter estimation error is derived, which accounts also for the presence of unknown but bounded disturbances 
and measurement noises. This latter result is of special interest in the context of {\em multi-model adaptive switching control} (MMASC) 
of uncertain systems \cite{Ba13}, \cite{morse97}-\nocite{hespanha,debruye,bald}\cite{BaBaTe13}, of which multi-model least-square parameter estimation constitutes one of the key elements. In this respect, it has been shown in  \cite{Ba13,BaBaTe13} that, in the case of a finite uncertainty set, by employing discerning
controllers it is possible to construct MMASC schemes which enjoy quite 
strong stability properties, namely exponential input-to-state stability. Hence, the results of the paper suggest that similar stability properties could be achieved also in the
case of continuously parameterized uncertainty. This issue will be the subject of further research.

The remainder of the paper is organized as follows. In Section II, we describe the framework under 
consideration. In Section III,  the main results of the paper are given concerning the existence and genericity of switching controllers ensuring parameter identifiability.
Section IV analyzes the properties of multi-model least-squares parameter estimation. 
Examples
are finally given in Section V. 

For the sake of clarity, all the proofs are reported in the Appendix section.

\emph{Notation.} Before concluding this section, let us introduce some notations
and basic definitions. Given a vector $v \in \mathbb R^n$, $| v |$ denotes its Euclidean norm.
Given a symmetric, positive semi-definite matrix $\, P $, we denote by
$\, {\lambda}_{\rm min}(P) $ and $\, {\lambda}_{\rm max}(P) $ the minimum
and maximum eigenvalues of $P$, respectively. Given a matrix $\, M
$, $\, M^{\top} $ is its transpose and $\, |
M | = \left [ \lambda_{\rm max} (M^{\top} M) \right ]^{1/2} $ its spectral
norm. Given a measurable time function $v : \mathbb R^+ \rightarrow \mathbb R^n $ and a time interval $\mathcal I \subseteq \mathbb R^+$,
we denote the $\mathcal L_2$ and $\mathcal L_\infty $ norms of $v(\cdot)$ on $\mathcal I$ as
$
\| v \|_{2,\mathcal I} = \sqrt{\int_{\mathcal I} | v (t)|^2 dt }
$
and
$
\| v \|_{\infty,\mathcal I} = \mbox{ess sup}_{t \in \mathcal I} {| v (t)|}
$
respectively. When $\mathcal I = R^+$, we simply write $\| v \|_{2}$ and $\| v \|_{\infty}$. 
Finally, we let $\mathcal L_2(\mathcal I)$ and $\mathcal L_\infty(\mathcal I) $ denote the sets of square integrable and, respectively, (essentially) bounded time functions on $\mathcal I$.

\section{Framework and objectives}\label{overview}

We consider a process described by an uncertain linear system  $\mathcal P(\theta)$
\begin{eqnarray} \label{equazione:sistema}
\left\{ \begin{array}{l} \dot x =
A(\theta) \, x + B(\theta) \, u \vspace{0.1cm} \\
y =   C(\theta) \, x \end{array} \right.\, 
\end{eqnarray}
where $x \in {\mathbb R}^{n_x}$ is the state, 
$u \in {\mathbb R}^{n_u}$ is the input, $y \in {\mathbb R}^{n_y}$ 
is the output, and $\theta \in {\mathbb R}^{n_\theta}$ is 
an unknown parameter vector belonging to the known
bounded set $\Theta \subseteq {\mathbb R}^{n_\theta}$.

The problem of interest is that of designing a controller $\mathcal C$
ensuring global discernibility, \emph{i.e.}, identifiability of 
the unknown parameter vector $\theta$ from observations of the plant input/output data 
\begin{eqnarray}
z = {\rm col}(u,y) \,. \nonumber
\end{eqnarray}
where {\em col} stands for column vector. 
In \cite{BaBaTe14}, the problem was addressed in the special case when the set $\Theta$ is finite and it was shown that, 
under mild assumption, global discernibility can be ensured by means of a LTI controller. When the set $\Theta$ is not finite, a single LTI controller
in general cannot ensure global discernibility by itself. Nevertheless, as will be shown in the following, it turns out that global discernibility can be achieved
by switching among a finite number of LTI controllers. Accordingly, let the controller be described by a switching linear system
\begin{eqnarray}
\left\{ \begin{array}{l} 
\dot \xi = F_{\sigma} \, \xi - G_{\sigma} \, y \vspace{0.1cm} \\
u = H_{\sigma} \, \xi - K_{\sigma} \, y
\end{array} \right.
\label{equazione:controllore}
\end{eqnarray}%
where $\xi \in {\mathbb R}^{n_{\xi}}$ is the controller state and
$\sigma:{\mathbb R}_+ \mapsto {\mathcal N}
:= \left\{ 1,2,\ldots, N\right\}$ is the 
switching signal, \emph{i.e.} the signal 
(right continuous) which identifies 
the index of the active system at each instant of time. Hereafter, it will be supposed that the switching signal $\sigma$
is generated so as to have a finite number of discontinuity points in every finite time interval.
For any $i \in \mathcal N$, $F_i$, $G_i$, $H_i$, and $K_i$ are constant matrices of appropriate dimensions.
In the sequel, we shall denote 
by $\mathcal C_i$ the LTI system with state-space representation $\{F_i,G_i,H_i,K_i\}$ and by $\mathcal C_\sigma$
the control law associated with the switching signal $\sigma$.

Denote by $\chi = \textrm{col} (x,\xi)$ and $z = \textrm{col} (u,y)$
the state and the output of the closed-loop system $(\mathcal P(\theta) / \mathcal C_\sigma)$ resulting from 
the interconnection of (\ref{equazione:sistema}) and (\ref{equazione:controllore})
when the unknown parameter takes value $\theta$ and the controller switching signal is $\sigma$.
The corresponding dynamics can be therefore expressed as
{\setlength\arraycolsep{1pt}
\label{}
\begin{eqnarray}\label{eq:closed-loop}
\left\{ \begin{array}{l} 
\dot \chi = \Psi_{\sigma}(\theta)  \, \chi  \\
z= \Lambda_{\sigma}(\theta) \chi
\end{array} \right.
\end{eqnarray}}%
where, for any $i \in \mathcal N$ and $\theta \in \Theta$,
\begin{eqnarray}
&& \Psi_{i}(\theta) = \left [ \begin{array}{cc} A(\theta) - B(\theta)\,K_i\,C(\theta)
& B(\theta) \, H_i \\ -G_i \, C(\theta) & F_i
\end{array} \right ], \quad \\
&& \Lambda_{i}(\theta) = \left [
\begin{array}{cc} -K_i\, C(\theta) & H_i
\\  C(\theta) & 0 \end{array} \right ] .
\nonumber
\end{eqnarray} 
Finally, let 
$
z(t,t_0,x_0,\xi_{0},\theta,\sigma)
$
denote the value at time $t$ of $z$ 
when the plant initial state at time $t_0$ is $x_{0}$, 
the controller initial state is $\xi_{0}$, the unknown parameter vector takes value $\theta$, and the controller switching signal is $\sigma$.
The following notions 
can be introduced. 

\begin{definition} \label{def:disc:CLD} 
Let the process be as in (\ref{equazione:sistema}) and assume that $u$ and $y$ are available for measurements.
Further, consider two distinct parameter vectors $\theta, \theta' \in \Theta$.
A switching controller of the form (\ref{equazione:controllore}) 
is said to be ($\theta$,$\theta'$)-\emph{discerning} if, 
for any time-interval $\mathcal I:= [t_0,\, t_0+T)$, with $T>0$, 
there exists a switching signal $\sigma: \mathcal I \rightarrow \mathcal N $ such that
\begin{eqnarray} \label{eq:main}
\left \| z(\cdot,t_0,x_0,\xi_{0},\theta,\sigma) - z(\cdot,t_0,x'_0,\xi'_{0},\theta',\sigma)  \right \|_{2, \mathcal I} \,\ne\, 0
\end{eqnarray}
for all nonzero quadruples of vectors
$(x_0,\xi_{0},x'_0,\xi'_{0})$. 
%
In addition, the switching controller (\ref{equazione:controllore}) is said to be 
\emph{globally discerning} if it satisfies condition (\ref{eq:main}) for all 
pairs $(\theta,\theta')$ of different parameter vectors.
\qedp
\end{definition} 
\salt

In words, when the switching controller (\ref{equazione:controllore}) is ($\theta$,$\theta'$)-discerning and
a discerning switching signal $\sigma$ is adopted, 
then  $\mathcal P(\theta)$ and $\mathcal P(\theta')$
cannot give rise to the same observation data 
when $\mathcal C_\sigma$ is in the feedback loop (unless the initial conditions are null). As a consequence, under global discernibility, 
it is possible to uniquely identify the unknown parameter vector $\theta$ by observing $z$ on the interval $\mathcal I$.

\section{Main results}\label{main}

In this section, we derive sufficient conditions for a switching control to be discerning and we show that, under mild assumptions, almost every switching controller
is discerning provided that the number $N$ of controller modes is sufficiently large.

To this end, notice preliminarily that a necessary condition for the existence of a discerning controller is that all the pairs $(A(\theta),C(\theta))$
are observable. In fact, the presence of unobservable dynamics would entail the existence
of non-zero trajectories of the closed-loop state $\chi$ corresponding to zero trajectories of the closed-loop output  $z$ and, hence, for which
it would be impossible to infer the plant mode. 
Accordingly, the following assumption is considered. 

\begin{enumerate}
\item[\bf A1.] The pair $(A(\theta),C(\theta))$ is observable for all $\theta \in \Theta$.
\end{enumerate} 

Let now $\varphi_i(s,\theta)$ denote the characteristic polynomial of the closed-loop system $(\mathcal P(\theta) / \mathcal C_i)$ resulting from the
feedback interconnection of the plant $\mathcal P(\theta)$ with the $i$-th controller mode $\mathcal C_i$. The following result holds. \salt

\begin{lemma} \label{lem:poly} Let assumption \textbf{A1} hold and suppose that, for any pair of distinct parameter vectors $\theta, \theta' \in \Theta$, there exist at least one index $i \in \mathcal N$ 
such that the two closed-loop characteristic polynomials $\varphi_i(s,\theta)$ and $\varphi_i(s,\theta')$ are coprime. Then, the following properties are true.
\begin{enumerate}[(i)]
\item the switching controller  (\ref{equazione:controllore})  is globally discerning;
\item condition (\ref{eq:main}) holds for any switching signal $\sigma$ such that each controller mode $i \in \mathcal N$ is active, at least, on an interval $\mathcal I_i \subset \mathcal I$
of positive measure.
\end{enumerate}
\qedp
\end{lemma} \salt

Let now $\bar n_\xi$ denotes the total number of elements of the controller matrices $(F_i, G_i, H_i, K_i)$ when the controller
order is $n_\xi$, and let $\mathcal D(\theta,\theta') \subseteq \mathbb R^{\bar n_\xi}$  be the set of  controller matrices\footnote{Here, with a little abuse
of notation, we identify the quadruples $(F_i, G_i, H_i, K_i)$ with an $\bar n_\xi$-dimensional vector containing all the elements of the matrices
$(F_i, G_i, H_i, K_i)$  according to a given order.} $(F_i, G_i, H_i, K_i)$ for which 
the two closed-loop polynomials $\varphi_i(s,\theta)$ and $\varphi_i(s,\theta')$ are coprime. Further, for a given number $N$ of controller modes, let
$\mathcal D_N \subseteq \mathbb R^{N \bar n_\xi}$ be the set of switching controllers  $(F_i, G_i, H_i, K_i), i  \in \mathcal N$ satisfying the hypothesis of Lemma 
\ref{lem:poly} (\emph{i.e.}, such that  for any pair of distinct parameter vectors $\theta, \theta' \in \Theta$, there exist at least one index $i \in \mathcal N$ 
such that the two closed-loop characteristic polynomials $\varphi_i(s,\theta)$ and $\varphi_i(s,\theta')$ are coprime). Since, in view of Lemma 
\ref{lem:poly}, all switching controllers belonging to $\mathcal D_N$ are globally discerning, we are now interested in studying the properties of the sets
$\mathcal D(\theta,\theta') $ and $\mathcal D_N$.

To this end, recall that coprimeness of the two polynomials $\varphi_i(s,\theta)$ and $\varphi_i(s,\theta')$ is equivalent to the fact that their 
\emph{Sylvester resultant} $R_i(\theta,\theta')$ (\emph{i.e.}, the determinant
of the Sylvester matrix associated with the two polynomials) is different from $0$. With this respect, we note that $R_i(\theta,\theta')$ depends polynomially on the elements
of the controller matrices $(F_i, G_i, H_i, K_i)$. Hence, when only a single pair of distinct parameter vectors $\theta, \theta' \in \Theta$ is taken into account,
the set of controller matrices $(F_i, G_i, H_i, K_i)$ for which $R_i(\theta,\theta')=0$, which is the complement of $\mathcal D(\theta,\theta') $, is an algebraic set.
Then, as well known, only two situations may occur: either $R_i(\theta,\theta')=0$ is always satisfied; or $R_i(\theta,\theta')=0$ is satisfied on a set with zero Lebesgue measure.
In this latter case, the set $\mathcal D(\theta,\theta') $ is \emph{generic}
\footnote{Recall that a subset $\mathcal X$ of a topological space is generic when it is open and dense: for any $x \in \mathcal X$, then there exists a neighborhood of $x$ contained in $\mathcal X$; for any $x \notin \mathcal X$, then every neighborhood of $x$ contains an element of $\mathcal X$.} (since it is the complement of a proper algebraic set). The following lemma, proved in \cite{Ba13}, provides sufficient conditions for such a favorable situation to occur.
 \salt

\begin{lemma} \label{lem:autp}
\cite{Ba13} Let assumption \textbf{A1} hold and consider two distinct parameter vectors $\theta, \theta' \in \Theta$. Then a  controller $(F_i, G_i, H_i, K_i)$ ensuring 
coprimeness of the two polynomials $\varphi_i(s,\theta)$ and $\varphi_i(s,\theta')$ exists if and only if 
the following two conditions hold:
\begin{enumerate}[(a)]
\item the transfer functions of $\mathcal P(\theta)$ and $\mathcal P(\theta')$
are different;
\item the characteristic polynomials of the uncontrollable parts of $\mathcal P(\theta)$ and $\mathcal P(\theta')$ are coprime.
\end{enumerate}
In addition, when conditions (a)-(b) holds, for any given controller order $n_\xi$ the set $\mathcal D(\theta,\theta')$ is generic and of full measure on $\mathbb R^{\bar n_\xi}$. 
\qedp
\end{lemma}  

Building on the above lemmas, under suitable regularity assumptions for the functions $A(\theta), B(\theta), C(\theta)$, it is possible to derive conditions for the existence of a global
discerning switching controller on the whole uncertainty set $\Theta$. In particular,  by exploiting the results of \cite{Sontag}, the following theorem can be stated. \salt

\begin{theorem} \label{thm1}
Let the uncertainty set $\Theta$ be contained in an analytic manifold $\mathcal M \subset \mathbb R^{n_\theta}$ of dimension $M$ and let the elements of the system matrices 
$A(\theta), B(\theta), C(\theta)$ be analytic functions of $\theta$ on $\mathcal M$. Further, let assumption \textbf{A1} and conditions (a)-(b) of Lemma \ref{lem:autp} hold on $\mathcal M$.
Then, provided that $N \ge 2 \, M + 1$, the set $\mathcal D_N$ is generic and of full measure on $\mathbb R^{N \, \bar n_\xi}$.
\qedp
\end{theorem} \salt

A few remarks are in order. First of all, notice that Theorem \ref{thm1} provides a bound on the number $N$ of controller modes
that may be needed in order to ensure identifiability of the unknown parameter $\theta$ in $\Theta$. Such a bound is consistent with the results of \cite{Ba13,BaBaTe14} where it is shown
that when the set $\Theta$ is finite, i.e., it is a $0$-dimensional manifold, one single LTI controller is generically discerning. As discussed in  \cite{Sontag}, the bound
$N \ge 2 \, M + 1$ for parameter distinguishability is tight for general maps (in the sense that when $N < 2 \, M + 1$ one can find counterexamples). However,
for specific cases, fewer controller modes can be sufficient. For instance, when $\theta$ is a scalar parameter, one can consider a single LTI controller $(F_i, G_i, H_i, K_i)$ 
and plot the root locus of the closed loop polynomial $\varphi_i(s,\theta)$ as a function of $\theta$. Then, global discernibility is guaranteed provided that such a root locus never cross itself.

Notice finally that the set of analytic functions considered in the statement of Theorem \ref{thm1} is quite general as it captures many function classes of interest 
(\emph{e.g.}, polynomials, 
trigonometric functions, exponentials, and also rational functions as long as away from singularities). 

\subsection{Accounting for additional control objectives}

In the foregoing analysis, only the global discernibility objective has been taken into account. However
generally speaking, a control law should typically satisfy other control objectives, the most fundamental one being stability.
With this respect, suppose that we want the switching controller to ensure closed-loop stability in a given subset $\bar \Theta$ of $\Theta$ together with
global discernibility. For instance, $\bar \Theta $ can represent the neighborhood of the nominal operating condition and the switching controller should
be designed so as to ensure: a satisfactory behavior in nominal conditions as well as the possibility of promptly identifying any departure from the nominal behavior (e.g, 
for fault-detection and isolation or for control reconfiguration purposes). An extreme case is when $\bar \Theta = \Theta$ so that we want to design a robust and globally discerning 
switching controller ensuring stability for any possible operating condition (of course this may be possible or not depending on the of size of the uncertainty set $\Theta$).

As well known, a sufficient condition to ensure stability under switching is the existence of a common Lyapunov function. For example, if we consider a quadratic parameter-dependent
Lyapunov function
$v(\chi) = \chi^\top \Pi(\theta) \chi$, then in order for the closed-loop system $(\mathcal P (\theta), \mathcal C_\sigma)$ to be stable for any $\theta \in \bar \Theta$ it is sufficient that
there exists $\Pi(\theta) = \Pi(\theta)^\top$ such that
\begin{eqnarray}
&& \Pi (\theta) \succ 0  \, , \label{eq:stab1} \\
&& \Psi_i (\theta)^\top \Pi (\theta) + \Pi(\theta) \Psi_i (\theta) \prec 0 \, ,  \label{eq:stab2}
\end{eqnarray}
for any $i \in \mathcal N$ and for any $\theta \in \bar \Theta$. When the set $\Theta$ is compact, by means of simple continuity arguments, it is immediate to show
that, for any given smooth $\Pi(\theta)$, the set of controller matrices $(F_i, G_i, H_i, K_i)$ satisfying (\ref{eq:stab2}) is an open subset of $\mathbb R^{\bar n_\xi}$.
Then, recalling that, by definition, any non-empty open set contains a closed-ball of positive radius, the following result on the existence of global discerning controllers ensuring also stability
can be readily stated. 

\begin{theorem} \label{thm2}
Let the same hypotheses of Theorem \ref{thm1} hold. Further, let the set $\bar \Theta$ be compact and suppose that there exists at least
one controller $(F_i, G_i, H_i, K_i)$ for which conditions (\ref{eq:stab1}) and (\ref{eq:stab2}) are satisfied with the Lyapunov matrix $\Pi (\theta)$
depending continuously on $\theta$.
Then, whenever $N \ge 2 \, M +1$, the set of switching controllers  $(F_i, G_i, H_i, K_i), i  \in \mathcal N$ that jointly satisfies (\ref{eq:stab1}) and (\ref{eq:stab2}) and ensures global discernibility
is non-negligible, in the sense that it contains a ball of positive radius in 
$\mathbb R^{N \, \bar n_\xi}$. \qedp 
\end{theorem}

\section{Multi-model least-squares parameter estimation}

In this section, we discuss how the unknown parameter vector $\theta$ can be estimated from the closed-loop data $z$ on an interval
$\mathcal I = [t_0, t_0+T]$ and we show that, when the data result from application of a discerning switching controller, the resulting estimate enjoys some nice 
properties even in the presence of unknown disturbances and measurement noises.

To this end, recall that, when the unknown parameter vector takes value $\theta$, the evolution of $z$ on the interval
$\mathcal I$ takes the form $z(t,t_0,x_0,\xi_0,\theta,\sigma)$. Then the set $\mathcal S_\sigma (\theta)$ of all possible
closed-loop data on the interval $\mathcal I$ associated with $\theta$ and with the switching signal $\sigma$ can be written as
{\setlength\arraycolsep{0pt} 
\begin{eqnarray*}
&& \mathcal S_\sigma (\theta) = \bigg \{  \hat z \in \mathcal L_2 (\mathcal I): \hat z(\cdot) = z(\cdot,t_0,\hat x_0, \hat \xi_0,\theta,\sigma) 
\mbox{ on } \mathcal I \\ && \quad \quad \quad \quad \qquad \quad \mbox{ for some } \hat x_0 \in \mathbb R^{n_x} , \, \hat \xi_0  \in \mathbb R^{n_\xi} \, \bigg \} .
\end{eqnarray*}}%
Hence a natural approach for estimating the plant unknown parameters is the {\em least-squares} one, which
amounts to selecting the parameter vector $\theta$ for which the distance between the observed close-loop data $z$ on the interval 
$\mathcal I$ and the set $\mathcal S_\sigma (\theta)$
is minimal. Accordingly, the optimal least-squares estimate $\hat \theta^\circ$ can be obtained as
\begin{equation}\label{eq:MDC}
\hat \theta^\circ  \in \arg \min_{\hat \theta \in \Theta}{\delta_\sigma(z,\hat \theta)} \, ;
\end{equation}
where
\begin{equation}
\delta_\sigma(z,\hat \theta)
 = \min_{\hat x_0 \in \mathbb R^{n_x}, \, \hat \xi_0 \in \mathbb R^{n_\xi}} \left \| z(\cdot) -  z(\cdot,t_0,\hat x_0, \hat \xi_0,\hat \theta,\sigma)  \right \|_{2,\mathcal I} \, .  
 \label{eq:distance}
\end{equation}

Concerning the computation of the distance (\ref{eq:distance}), for any possible feedback loop
$(\mathcal P (\theta), \mathcal C_\sigma)$, let  $\Phi_\sigma(t,t_0,\theta)$ denote its state transition matrix  and let $W_\sigma (\theta)$ be 
its observability Gramian on the interval $\mathcal I$, \emph{i.e.},
\[
W_\sigma (\theta) = \int_{\mathcal I} \Phi_\sigma(t,t_0,\theta)^{\top} \Lambda_{\sigma(t)} (\theta)^\top \Lambda_{\sigma(t)} (\theta) \Phi_\sigma(t,t_0,\theta) dt \, .
\]
Notice that, for any globally discerning switching control law, the observability Gramian 
$W_\sigma (\theta)$ turns out to be
positive definite for any $\theta \in \Theta$ (otherwise there would be zero output trajectories corresponding to non-zero state trajectories and
parameter identification would not be possible). Hence, in this case, 
the minimization in (\ref{eq:distance}) yields
{\setlength\arraycolsep{0pt} 
\begin{eqnarray*}
&& {\delta_{\sigma}(z,{\hat \theta})} = \bigg ( \int_{\mathcal I} \bigg | z(t) - \Lambda_{\sigma(t)} (\hat \theta) \Phi_{\sigma} (t,t_0,\hat \theta) \left ( W_\sigma (\hat \theta)  \right )^{-1} 
\\ && \quad  {} \times \int_{\mathcal I}  \Phi_{\sigma} (\tau,t_0,\hat \theta)
^\top \, \Lambda_{\sigma(\tau)}  (\hat \theta)^\top z(\tau) \, d\tau \,  \bigg |^2 d t \, \bigg )^{1/2}.
\end{eqnarray*}}%
For further considerations on how this quantity can be computed in practice the interested reader is referred to Appendix A of \cite{BaBaTe13},
where a similar problem is addressed.

From Definition 1, it is immediately clear that when a globally discerning switching control law is adopted, for any non null $z$ the $\delta_\sigma(z,\hat \theta)$ is zero if and only if
$\hat \theta$ coincides with $\theta$, the true parameter vector.\salt

\begin{proposition}\label{prop:least-squares}
Let the switching controller (\ref{equazione:controllore}) be globally discerning and
a discerning switching signal $\sigma$ be adopted on the observation interval $\mathcal I$. Further, let the observed
data $z$ be generated by the closed-loop system (\ref{eq:closed-loop}) from initial condition 
$\chi_0 = (x_0,\xi_0) \ne 0 $. Then, $\hat \theta^\circ = \theta$.
\qedp
\end{proposition}
\salt

While the above proposition illustrates the theoretical effectiveness of the leas-squares estimation criterion in ideal conditions, in practice computation of the minimum
in (\ref{eq:MDC}) can be a quite challenging task when the set $\Theta$ is not finite. In this case, a standard approach is the {\em multi-model} one which amounts to considering only a finite
number, say L, of possible parameter values by constructing the finite set 
$\Theta_L =\{   \theta_\ell, \, \ell = 1, \ldots, L  \} \subseteq \Theta$. 
Typically, $\Theta_L$ is obtained by sampling the set $\Theta$ with a given guaranteed density $\varepsilon$, so that for any $\theta \in \Theta$ there exists
at least one $\theta_\ell \in \Theta_L$ such that $| \theta - \theta_\ell | \le \epsilon$. When such a condition is satisfied, we say that $\Theta_L$ is {\em $\epsilon$-dense} in $\Theta$.
Accordingly, the following {\em multi-model least squares criterion}  can be used to estimate the unknown parameter vector $\theta$
\begin{equation}\label{eq:MDC:MM}
\hat \theta_L  \in \arg \min_{\hat \theta \in \Theta_L}{\delta_\sigma(z,\hat \theta)}
\end{equation}
as an alternative to (\ref{eq:MDC}). 

\begin{remark}
Guidelines on how to choose an {\em $\epsilon$-dense} finite covering for $\Theta$
can be found, for instance, in \cite{debruye2}-\nocite{baldi2}\cite{Buchstaller}. \qedp
\end{remark}

\subsection{Properties of multi-model least-squares parameter estimation}

When analyzing of the properties of an estimation criterion, either optimal as in (\ref{eq:MDC}) or approximate as in (\ref{eq:MDC:MM}),
it is important to take into account also the effects of process disturbances and measurement noises. With this respect, in the following analysis we suppose that the
plant state and measurement equations are
affected by additive disturbances $d$ and $n$,
 respectively, \emph{i.e.},
\begin{equation}
\mathcal P(\theta) : \left \{ \begin{array}{rcl} \dot x & =
& A(\theta) \, x + B(\theta) \,
u + d \\
y & = &  C(\theta) \, x + n
\end{array} \right . \label{equazione:sistema:rumore}
\end{equation}
with $d \in \mathbb R^{n_x}$ and $n \in \mathbb R^{n_y}$.
Then, it is an easy matter to verify that a state space
representation of the closed-loop system 
$(\mathcal P(\theta)/\mathcal C_\sigma) $ takes the form
\begin{equation}
 \left \{
\begin{array}{rcl} \dot \chi & = & \Psi_{\sigma}(\theta) \, \chi + \Xi_{\sigma} (\theta) \, v  \\
z & = &  \Lambda_{\sigma} (\theta) \, \chi +
\Gamma_{\sigma} \, v
\end{array} \right . \label{equazione:feedback:rumore}
\end{equation}
where $v = (d,n)$ and
\[
 \Xi_{i} (\theta)  = \left [ \begin{array}{cc} I & B(\theta) \, K_i
\\ 0 & G_i
\end{array} \right ] \, , \quad \Gamma_i = \left [
\begin{array}{cc} 0 & K_i  
\\  0 & I \end{array} \right ] \, ,
\]
for any $i \in {\mathcal N} $ and $\theta \in \Theta$.
Further, thanks to linearity, the
closed-loop data $z$ can be decomposed as
\begin{equation} \label{eq:decomposition}
z = z^{({\rm n})}  + z^{({\rm f})}
\end{equation}
where $z^{({\rm f})}$ is the forced response and $z^{({\rm n})} (t)$ is the natural response
which can be written as
\[
z^{({\rm n})} (t) =  z(t,t_0, x_0, \xi_0, \theta,\sigma) 
\]
with $z(t,t_0, x_0, \xi_0, \theta,\sigma)$  the same function of the previous sections.
Notice also that the forced response $z^{({\rm f})}$ can be bounded in terms of the disturbance amplitude as follows. 

\begin{proposition}
Let the set $\Theta$ be compact and let the elements of $A(\theta)$, $B(\theta)$, and $C(\theta)$ depend continuously on $\theta$.
Then, for any $\mathcal I = [t_0, t_0 + T] $, there exists a positive real $\gamma$ such that
\begin{equation}\label{eq:bound2}
\| z^{({\rm f})} \|_{2, \mathcal I} \le \gamma \, \| v \|_{\infty,\mathcal I}  \, .
\end{equation}
\qedp
\end{proposition}

Notice now that, by virtue of the triangular inequality, we have
\begin{equation}\label{eq:triangular}
 {\delta_{\sigma}(z^{({\rm n})} ,{\hat \theta})} - \| z^{({\rm f})} \|_{2, \mathcal I}  \le {\delta_{\sigma}(z,{\hat \theta})} \le {\delta_{\sigma}(z^{({\rm n})} ,{\hat \theta})} + \| z^{({\rm f})} \|_{2, \mathcal I} 
\end{equation}
for any $\hat \theta \in \Theta$. Hence, the properties of the least-squares estimate $\hat \theta_L$ can be investigated by deriving bounds on $ {\delta_{\sigma}(z^{({\rm n})} ,{\hat \theta})}$.
In this respect, the following result is relevant. 

\begin{proposition}
Consider the same assumptions as in Proposition \ref{prop:least-squares}. Then, for any 
$\hat \theta \in \Theta$,
\begin{eqnarray*}
\lefteqn{ [{\delta_{\sigma}(z^{({\rm n})} ,{\hat \theta})}]^2} \\ && {} = \left [ \begin{array}{r} \chi_0 \\ -V_\sigma(\theta,\hat \theta) \chi_0 \end{array} \right ]^\top  
  W_\sigma(\theta,\hat \theta) \left [ \begin{array}{r} \chi_0 \\ -V_\sigma(\theta,\hat \theta) \chi_0 \end{array} \right ]
\end{eqnarray*}
where
\begin{eqnarray}
U_\sigma(\theta,\hat \theta) &=& 
 \int_{\mathcal I}  \Phi_{\sigma} (\tau,t_0,\hat \theta)
^\top \, \Lambda_{\sigma(\tau)}  (\hat \theta)^\top   \nonumber \\
&& \times \Lambda_{\sigma(\tau)}  ( \theta)  \, \Phi_{\sigma} (\tau,t_0,\theta) \, d\tau \\
V_\sigma(\theta,\hat \theta) &=& \left ( W_\sigma (\hat \theta)  \right )^{-1}  U_\sigma(\theta,\hat \theta)   \label{eq:V}  \\
W_\sigma(\theta,\hat \theta) &=& \left [ \begin{array}{ll} W_\sigma ( \theta) & U_\sigma(\theta,\hat \theta)^\top   \\ U_\sigma(\theta,\hat \theta)   & W_\sigma (\hat \theta)  \end{array} \right] \, .  \label{eq:W}
\end{eqnarray}
\qedp
\end{proposition}

For any pair of parameter vectors $\theta,\hat \theta \in \Theta$, the {\em joint observability Gramian} $W_\sigma(\theta,\hat \theta)$ provides information concerning
the degree of distinguishability between the two closed-loop systems $(\mathcal P(\theta), \mathcal C_\sigma )$ and $(\mathcal P(\hat \theta), \mathcal C_\sigma)$. In fact,
whenever the switching control law $\mathcal C_\sigma$ is globally discerning, the matrix $W_\sigma(\theta,\hat \theta)$ is singular if and only if $\theta = \hat \theta$
(this is a straightforward consequence of Proposition \ref{prop:least-squares}). Then,
we can derive the following result. \salt

\begin{lemma}\label{lemma:K-function}
Let the same assumptions as in Proposition \ref{prop:least-squares} hold. Further, let the set $\Theta$ be compact and let the elements of $A(\theta)$, $B(\theta)$, and $C(\theta)$ depend continuously on $\theta$. 
Moreover, let the switching controller (\ref{equazione:controllore}) be globally discerning and
a discerning switching signal $\sigma$ be adopted on the observation interval $\mathcal I$. Then, there exist two class $\mathcal K$ functions\footnote{Recall
that a function $\varphi: \mathbb R^+ \rightarrow \mathbb R^+$ belongs to class $\mathcal K$ if it is continuous, strictly increasing, and $\varphi(0) = 0$.} $\alpha (\cdot)$ 
and $\beta (\cdot)$ such that
\begin{equation} \label{eq:K-function}
\alpha(|\theta-\hat \theta|) \, | \chi_0 | \le {\delta_{\sigma}(z^{({\rm n})} ,{\hat \theta})} \le \beta(|\theta-\hat \theta|) \, | \chi_0 |
\end{equation}
for any $\hat \theta \in \Theta$.
\qedp
\end{lemma}\salt

The importance of Lemma \ref{lemma:K-function} is that it allows to bound the distance ${\delta_{\sigma}(z^{({\rm n})} ,{\hat \theta})}$, pertaining to the noise-free dynamics,
in terms of the initial state $\chi_0$ and of the distance between the true parameter vector $\theta$ and the candidate estimate $\hat \theta$. In particular, the left-most inequality in
(\ref{eq:K-function}) ensures that ${\delta_{\sigma}(z^{({\rm n})} ,{\hat \theta})}$ cannot be small when the discrepancy $\theta-\hat \theta$ is large, whereas the right-most inequality
ensures that ${\delta_{\sigma}(z^{({\rm n})} ,{\hat \theta})}$ nicely degrades to $0$ as the estimate $\hat \theta$ approaches the true value $\theta$. With respect to the latter observation,
from (\ref{eq:K-function}) it follows that when  $\Theta_L$ is $\epsilon$-dense in $\Theta$ there always exists a parameter vector $\theta_\ell \in \Theta_L$ such that
${\delta_{\sigma}(z^{({\rm n})} ,{\theta_\ell})} \le \beta(\epsilon) \, | \chi_0 |$. By exploiting inequalities (\ref{eq:triangular}) and (\ref{eq:K-function}), the main result of this section can finally be stated.
\salt

\begin{theorem}
Let the same assumptions as in Lemma \ref{lemma:K-function} hold. Further, let $\Theta_L$ be $\epsilon$-dense in $\Theta$. Then, the estimate $\hat \theta_L$ obtained as in
(\ref{eq:MDC:MM}) is such that
\begin{equation}\label{eq:bound}
|\theta-\hat \theta_L | \le \alpha^{-1} \left ( \beta(\varepsilon) + \frac{2 \, \gamma \, \| v \|_{\infty,\mathcal I} }{| \chi_0 |} \right )
\end{equation}
where $\alpha^{-1}(\cdot)$ is the inverse of $\alpha(\cdot)$. \qedp
\end{theorem} 

Concerning the upper bound on the estimation error provided by inequality (\ref{eq:bound}), it can be seen that  the term $\beta(\varepsilon)$ 
(which decreases the denser the sampling $\Theta_L$ is) accounts
for the fact that only a finite number of models is considered, while the term
$2 \, \gamma \, \| v \|_{\infty,\mathcal I} / |\chi_0 | $ can be seen as a sort of noise-to-signal ratio, and indeed goes to $0$ as the disturbance $\| v \|_{\infty,\mathcal I}$ goes to $0$. 

\section{An example}

In the following, a simple example is provided in order to illustrate how identifiability of an uncertain parameter vector
can be achieved my means of switching control. To this end, consider an LTI plant with system matrices
\[
A(\theta) = \left [ \begin{array}{rr} 0 & 1 \\ 0 & -a \end {array} \right ] , \quad B(\theta) = \left [ \begin{array}{r}  0 \\ b \end {array} \right ]
, \quad C(\theta) = \left [ 1 \quad 0 \right ],
\]
with $\theta = (a , b)$, and let the switching controller be a purely proportional one
\[
u = - K_\sigma \, y \, .
\] 
Since, the plant transfer matrix is $P(s,\theta) = b / [s \, (s + a)]$, each closed-loop characteristic polynomial takes the form
$
\varphi_i (s,\theta) = s^2 + a \, s + b \, K_i \, .
$

\subsection{One uncertain parameter} 

Suppose first, for illustration purpose, that only the gain $b$ is uncertain whereas
$a$ is perfectly known, i.e., $\Theta = \left \{ (a,b): a=a_0, b \in [b_1 , b_2]  \right \}$. Notice that assumption \textbf{A1} holds whenever $a_0 \ne 0$. Hence, in this case, we can exploit Lemma 1 and consider, for any pair $b, b'$ the two polynomials
\begin{eqnarray*}
&& \varphi_i (s,\theta) = s^2 + a_0 \, s + b \, K_i \, , \\
&& \varphi_i (s,\theta') = s^2 + a_0 \, s + b' \, K_i \, .
\end{eqnarray*}
As it can be easily verified, the Sylvester resultant of such polynomials is
\[
R_i (\theta , \theta') = K_i^2 \, (b-b')^2 .
\]
Then, it can be seen that if $K_i \ne 0$, the resultant is different from $0$ whenever $b$ and $b'$ are different.  Hence, in this
case, a single proportional controller with non-null gain is globally discerning and there is no need for
considering a switching controller in that $\mathcal D_1 = \{ K_1 \ne 0 \}$. Similar considerations hold when $a$ is uncertain
and $b$ is perfectly known.

\subsection{Two uncertain parameters} 
Suppose now that both $a$ and $b$ are uncertain, \emph{i.e.},
$\Theta = \left \{ (a,b): a \in [a_1 , a_2], b \in [b_1 , b_2]  \right \}$. Again, assumption \textbf{A1} holds provided that $0 \notin [a_1 , a_2]$. Straightforward calculations allow to see that, in this case, the resultant of the two polynomials
\begin{eqnarray*}
&& \varphi_i (s,\theta) = s^2 + a \, s + b \, K_i \, , \\
&& \varphi_i (s,\theta') = s^2 + a' \, s + b' \, K_i \
\end{eqnarray*} 
is
\[
R_i (\theta , \theta') = K_i^2 \, (b-b')^2 - K_i (b \, a' - a \, b') (a -a').
\]
Hence, a single controller is not sufficient for global discernibility as by choosing
\begin{equation}\label{eq:ab}
(b-b')^2 = \varepsilon , \quad = K_i  \, \varepsilon
\end{equation}
one has $R_i (\theta , \theta') = 0$. In fact, since $\varepsilon$ can be arbitrarily small, it is always possible to find $a,b,a',b'$ so as to satisfy (\ref{eq:ab}) regardless of the amplitude of the uncertainty set $\Theta$. Then $\mathcal D_1 = \emptyset$. \\
On the contrary, it can be seen that a switching controller with two modes, $N=2$, is generically globally discerning.
To see this, notice that
\[
\left [ \begin{array}{l} R_1 (\theta , \theta') \\  R_2 (\theta , \theta') \end{array} \right ]
= \left [ \begin{array}{ll}  K_1^2 & - K_1 \\ K_2^2 & - K_2 \end{array} \right ] 
\left [ \begin{array}{l} (b-b')^2 \\  (b \, a' - a \, b') (a -a') \end{array} \right ] 
\]
and
\[
\det  \left [ \begin{array}{ll}  K_1^2 & - K_1 \\ K_2^2 & - K_2 \end{array} \right ] = K_1 \, K_2 (K_2 -K_1).
\]
If we choose $K_1$ and $K_2$ such that  $K_1 \ne K_2$, $K_1 \ne 0$, and $K_2 \ne 0$ the above determinant turns out to be
different from $0$. As a consequence, in this case, the two resultants $R_1 (\theta , \theta')$ and $R_2 (\theta , \theta')$ 
can simultaneously vanish if and only if $(b-b')^2 = 0$ and $ (b \, a' - a \, b') (a -a') = 0$ which is equivalent to
$a=a'$ and $b=b'$. Hence, we have that $\mathcal D_2 = \{ (K_1,K_2): \; K_1 \ne K_2, \, K_1 \ne 0, \, K_2 \ne 0 \}$ which is generic and of full measure in $\mathbb R^2$.

\section{Conclusions}

In this paper, we have addressed the problem of identifying the parameters of an uncertain 
linear system by means of switching control. 
It was shown that even when the uncertainty set is not finite, 
parameter identifiability can be generically ensured by 
switching among a finite number of linear time-invariant controllers. 
In particular, the results show that an upper bound on the number of controller modes needed for parameter identifiability can be given in terms of the dimension of the uncertainty set. 
The results also indicate that
the seemingly conflicting goals of ensuring parameter identifiability as well as a satisfactory behavior of the feedback system can be simultaneously accomplished by means of switching control. 

Several practical aspects have also been discussed. In particular, 
we have analyzed the properties of 
least-squares parameter estimation in connection with the use of discerning controllers,
providing bounds on the worst-case parameter estimation error
in the presence of: i) finite covering of the uncertainty set; and 
ii) bounded disturbances affecting the process dynamics as well as measurement noises. 
 
The results lend themselves to be extended in various directions.
Most notably, these results find a very natural application
in the context of switching control for uncertain systems. In this respect, 
we envision that the analysis tools introduced in this paper should 
lead to the development of novel control reconfiguration
algorithms capable of achieving input-to-state stability 
for uncertain systems even when the uncertainty set is described 
by a continuum.

\section*{Appendix}

{\em Proof of Lemma 1:} Consider two distinct parameter vectors $\theta, \theta' \in \Theta$ and consider an index $i$ for which $\varphi_i(s,\theta)$ and $\varphi_i(s,\theta')$
are coprime. Let the controller $\mathcal C_i$ be active on an interval $\mathcal I_i = [\underline t , \bar t] \subset \mathcal I$.
Consider now a nonzero quadruples of vectors $(x_0,\xi_0,x_0',\xi'_0)$ representing possible initial states of the two feedback loops $(\mathcal P (\theta) / \mathcal C_\sigma)$ 
and $(\mathcal P (\theta') / \mathcal C_\sigma)$ at time $t_0$. Let $(\underline x,\underline \xi, \underline x', \underline \xi')$ be the corresponding states 
that are reached at time $\underline t$, \emph{i.e.}, at the beginning of the time
interval $\mathcal I_i$, under the switching law $\sigma$. Suppose now the switching signal $\sigma$ is chosen so as to satisfy a dwell-time condition, \emph{i.e.}, in such a way that there exists a lower bound
$\tau_{\rm dwell}$ on the time interval between subsequent variations of the controller index. Then, it is immediate to see that, under such a switching law, 
when $(x_0,\xi_0,x_0',\xi'_0) \ne 0$ then also $(\underline x,\underline \xi, \underline x', \underline \xi') \ne 0$. In fact, such a state is reached after switching a finite number of times between 
autonomous linear systems, i.e., the feedback loops, and
it is known that an autonomous linear system cannot reach the zero state in finite time starting from a non-zero initial state. Notice now that, under assumption \textbf{A1}, coprimeness of
the polynomials $\varphi_i(s,\theta)$ and $\varphi_i(s,\theta')$ implies observability of the parallel system
\[
\left \{ \begin{split}
& \left [ \begin{array}{c} \dot \chi \\ \dot \chi' \end{array}  \right ] = \left [ \begin{array}{cc} \Psi_i (\theta) & 0 \\ 0 & \Psi_i (\theta)  \end{array}  \right ]
\left [ \begin{array}{c} \chi \\ \chi' \end{array}  \right ] \\
& \tilde z = \left[ \Lambda_i (\theta) \quad -  \Lambda_i (\theta) \right ] \left [  \begin{array}{c} \chi \\ \chi' \end{array}  \right ] 
\end{split}
\right .
\]
(see for instance Proposition 1 of \cite{Ba13}).
Then, if we initialize such a system as $ (\chi (\underline t) , \chi' (\underline t) ) = (\underline x,\underline \xi, \underline x', \underline \xi') \ne 0$ at time $\underline t$,
we have that $\tilde z$ is different from $0$ a.e. on $\mathcal I_i = [\underline t , \bar t]$,
where ``a.e.'' stands for ``almost everywhere'', \emph{i.e.} everywhere 
except on a set of zero Lebesgue measure. This, in turn, implies that
$z(t,\underline t, \underline x, \underline \xi, \theta, i) \ne z(t,\underline t, \underline x', \underline \xi', \theta', i)$, or equivalently, 
$z(t,t_0, x_0, \xi_0, \theta, \sigma) \ne z(t, t_0, x_0',  \xi_0', \theta', i)$,  a.e. on $\mathcal I_i = [\underline t , \bar t]$. Then, by choosing a switching signal $\sigma$ which satisfies
a dwell-time condition and is such that each controller mode $i$ is active, at least, on an interval $\mathcal I_i \subset \mathcal I$ of positive measure, the same line reasoning can be repeated
for any pair $\theta, \theta' \in \Theta$ with $\theta \ne \theta' $, thus concluding the proof. \qedp

{\em Proof of Theorem 1:} 
Notice first that the resultant $R_i(\theta,\theta')$ of the two polynomials $\varphi_i(\theta)$ and $\varphi_i(\theta')$ is a polynomial (and hence analytic) function of the elements
of the matrices $(A(\theta), B(\theta), C(\theta))$, $(A(\theta'), B(\theta'), C(\theta'))$, and $(F_i, G_i, H_i, K_i)$. This, in turn, implies that $R_i(\theta,\theta')$ is an analytic function of $\theta$ and
$\theta'$ (since the composition of analytic functions is analytic). Notice now that, under the stated hypotheses, Lemma 2 ensures that, for any pair $\theta, \theta' \in \mathcal M$ with $\theta \ne \theta'$,
it is possible to find at least one set of matrices $(F_i, G_i, H_i, K_i)$ such that $\varphi_i(\theta)$ and $\varphi_i(\theta')$ are coprime and, hence,
$R_i(\theta,\theta') \ne 0$. Recall, finally, that the set $\mathcal D_N$ corresponds to the set of switching controllers $(F_i,G_i,H_i,K_i) i \in \mathcal N$ for which the vector function
 ${\rm col} \left ( R_i(\theta,\theta') , i \in \mathcal N \right )$ is different from $0$ for any pair $\theta, \theta' \in \mathcal M$ with $\theta \ne \theta'$.
Then, proceeding as in the proof of Theorem 2 of \cite{Sontag}, we can conclude that $\mathcal D_N$ is generic and of full measure on $\mathbb R^{N \bar n_\xi}$ whenever $N \ge 2 \, M +1$.
\qedp

{\em Proof of Theorem 2:} 
Let $(F_i,G_i,H_i,K_i)$ be a controller for which conditions (\ref{eq:stab1}) and (\ref{eq:stab2}) are satisfied with $\Pi (\theta)$ continuous in $\theta$. 
Further, consider a closed ball $\mathcal B (\varepsilon)$ in the controller parameter space $\mathbb R^{\bar n_\xi}$ centered in $(F_i,G_i,H_i,K_i)$ and with radius $\varepsilon$
and let 
\[
\begin{split}
& \beta (\varepsilon)  = \\
& \max_{(F,G,H,K) \in \mathcal B (\varepsilon)} \max_{\theta \in \bar \Theta } \lambda_{\rm max} 
\left \{  \Psi_i (\theta)^\top \Pi (\theta) + \Pi(\theta) \Psi_i (\theta) \right \} \, .
\end{split}
\]
Note that, in view of the compactness of  $\bar \Theta$ and of the continuity of $\Psi (\theta)$ and $\Pi (\theta)$, we have that $\max_{\theta \in \bar \Theta } \lambda_{\rm max} 
\left \{  \Psi_i (\theta)^\top \Pi (\theta) + \Pi(\theta) \Psi_i (\theta) \right \} = \beta(0) < 0$. Moreover, under the considered hypotheses, it is easy to show that 
$\beta (\varepsilon) $ depends continuously on $\varepsilon$ (in this respect, notice that 
$\Psi_i (\theta)$ is an affine function of the controller matrices $(F_i,G_i,H_i,K_i)$).
Hence, this implies  the existence of $\varepsilon >0$ such that $\beta (\varepsilon) <0$, 
\emph{i.e.}, such that  all the controllers in $ \mathcal B (\varepsilon)$ satisfies 
(\ref{eq:stab2}) with the same Lyapunov matrix $\Pi (\theta)$. As a consequence, the set 
$\mathcal G \subset \mathbb R^{\bar n_\xi}$ of all controllers satisfying (\ref{eq:stab2})  with the 
Lyapunov matrix $\Pi (\theta)$ turns out to be open, and $\mathcal G^N$ will be open as well. Finally, when $N \ge 2 M +1$, the set $\mathcal D_N$ is generic and
of full measure on $\mathbb R^{N \, \bar n_\xi}$ and, hence, $\mathcal G^N \cap \mathcal D_N$ is non-negligible.
\qedp

{\em Proof of Proposition 1:} This  is a straightforward consequence of the fact that, when $ \hat \theta \ne \theta$, we cannot have 
$z(t) = z(t, t_0, \hat x_0, \hat \xi_0, \hat \theta , \sigma )$ a.e. on $\mathcal I$, since the 
observed closed-loop data are generated as $z(t) = z(t, t_0, x_0, \xi_0, \theta , \sigma )$ with $(x_0,\xi_0) \ne 0$
and the control law $\mathcal C_\sigma$ is supposed to be discerning. Hence, $\delta_\sigma(z,\hat \theta) > 0$ whenever
$\hat \theta \ne \theta$, and $\delta_\sigma(z,\theta) = 0$ since by hypothesis $z \in \mathcal S_\sigma (\theta)$.
\qedp

{\em Proof of Proposition 2:}
Recalling that the forced response $z^{({\rm f})}$ of the switching linear system $(\mathcal P (\theta) / \mathcal C_\sigma)$ can be written
as
\[
\begin{split}
&z^{({\rm f})} (t) =  \Lambda_{\sigma(t)} (\theta) \int_{t_0}^{t}  \Phi_\sigma(\tau,t_0,\theta) \, \Xi_{\sigma(\tau)} (\theta) \, v(\tau) \, d \tau \\
&  \quad \quad \quad + \Gamma_{\sigma(t)} \, v (t) \, ,
\end{split}
\]
it is an easy matter to see that 
\[
\begin{split}
&|z^{({\rm f})} (t)| \le \left | \Lambda_{\sigma(t)} (\theta) \right | \int_{t_0}^{t}  \left | \Phi_\sigma(\tau,t_0,\theta) \, \Xi_{\sigma(\tau)} (\theta) \right |   \, d \tau \, \|  v \|_{\infty, \mathcal I} \\
 &\quad \quad \quad + \left | \Gamma_{\sigma(t)} \right | \, \left | v (t) \right | \, . 
 \end{split}
 \]
From the latter inequality, the bound in (\ref{eq:bound2}) can be readily obtained since, by hypothesis, the switching signal $\sigma$ contains only a finite number of discontinuity points in $\mathcal I$.
\qedp

{\em Proof of Proposition 3:} It follows from standard calculations by replacing $z^{({\rm n})} (t)$ with $ \Lambda_{\sigma (t)} (\theta) \, \Phi_\sigma (t,t_0,\theta)$ in the expression for the distance $\delta_\sigma (z^{({\rm n})}, \hat \theta)$. \qedp

{\em Proof of Lemma 3:} 
In view of Proposition 3, we have that
\[
\delta_\sigma(z^{({\rm n})},\hat \theta) \le \beta_1 (\theta, \hat \theta) | \chi_0 |
\]
with
\[
\begin{split}
 &\beta_1^2 (\theta, \hat \theta)  = \\ & \quad \lambda_{\rm max} \left \{ \left [ \begin{array}{c} I \\ - V_\sigma (\theta, \hat \theta) \end{array} \right ]^\top
 W_\sigma (\theta, \hat \theta)  
 \left [ \begin{array}{c} I \\ - V_\sigma (\theta, \hat \theta) \end{array} \right ]  \right \}
\, .
\end{split}
\]
Notice that  $\beta_1 (\theta, \hat \theta) $ depends continuously on $\theta$ and $\hat \theta$ and, in addition, $\beta_1 (\theta, \hat \theta) = 0$
if and only if $\theta = \hat \theta$ since the switching law is supposed to be discerning. Then,
the class $\mathcal K$ function $\beta(\rho)$ can be taken equal to $\max_{\theta, \hat \theta \in \Theta, \, | \theta - \hat \theta| \le \rho , } \beta_1 (\theta, \theta) $.
As for the lower bound, notice that Proposition 3 implies also that
{\setlength\arraycolsep{1pt}
\begin{eqnarray*}
\left [ \delta_\sigma(z^{({\rm n})},\hat \theta) \right ]^2 &\ge& \lambda_{\rm min} \left \{ W_\sigma (\theta,\hat \theta) \right \} \, ( | \chi_ 0|^2 + | V_\sigma (\theta,\hat \theta) \chi_0|^2) 
\\ &\ge& \lambda_{\rm min} \left \{ W_\sigma (\theta,\hat \theta) \right \} \,  | \chi_ 0|^2 \, .
\end{eqnarray*}}%
Since $ \lambda_{\rm min} \left \{ W_\sigma (\theta,\hat \theta) \right \} $ depends continuously on $\theta,\hat \theta$ and is equal to $0$ if and only if
$\theta = \hat \theta$ (again thanks to the discernibility of the switching law), then a class $K$ function $\alpha (|\theta-\hat \theta|)$ can be found
that satisfies the inequality $\lambda_{\rm min} \left \{ W_\sigma (\theta,\hat \theta) \right \}  \ge \alpha^2 (|\theta-\hat \theta|) $ for any $\theta, \hat \theta \in \Theta$.
In particular, $\alpha (|\theta-\hat \theta|)$ can be constructed as in the proof of Theorem 2 of \cite{AlBaBaTAC} to which the reader is referred for additional details.
\qedp

{\em Proof of Theorem 3:}
Since $\Theta_L$ is $\epsilon$-dense in $\Theta$, there exists at least one $\hat \theta^* \in \Theta_L$ such that $| \theta -\hat \theta^*| \le \epsilon$. For such a
$\hat \theta^*$, one has
\begin{eqnarray*}
{\delta_{\sigma}(z,{\hat \theta^*})} &\le& {\delta_{\sigma}(z^{({\rm n})} ,{\hat \theta^*})} + \| z^{({\rm f})} \|_{2, \mathcal I} \\ 
&\le& \beta (| \theta - \hat \theta^*|) | \chi_0 | + \| z^{({\rm f})} \|_{2, \mathcal I}  \\ &\le& \beta (\epsilon) | \chi_0 | + \| z^{({\rm f})} \|_{2, \mathcal I}  \, .
\end{eqnarray*}
Since the estimate $\hat \theta_L$ is optimal in $\Theta_L$, one has also
\[
{\delta_{\sigma}(z,{\hat \theta_L})} \le  {\delta_{\sigma}(z,{\hat \theta^*})} \le \beta (\epsilon) | \chi_0 | + \| z^{({\rm f})} \|_{2, \mathcal I}  \, .
\]
Further, by exploiting the lower bound in Proposition 3, we can write
\begin{eqnarray*}
{\delta_{\sigma}(z,{\hat \theta_L})} &\ge& {\delta_{\sigma}(z^{({\rm n})} ,{\hat \theta_L})} - \| z^{({\rm f})} \|_{2, \mathcal I}  \\
&\ge & \alpha (| \theta - \hat \theta_L|) \, |\chi_0|  - \| z^{({\rm f})} \|_{2, \mathcal I} .
\end{eqnarray*}
Combining the two latter inequalities, we obtain
\[
 \alpha (| \theta - \hat \theta_L|) \, |\chi_0| \le  \beta (\epsilon) | \chi_0 | + 2 \, \| z^{({\rm f})} \|_{2, \mathcal I} 
\]
which can be written as (\ref{eq:bound}), Proposition 2 and the fact that any class $\mathcal K$ function is invertible. 
\qedp

\bibliographystyle{IEEEtran} 
      
\bibliography{discerning}

\end{document}